\documentclass[12pt,a4paper]{article}

\usepackage{epsfig}

\newcommand{\ie}{{\it i.e. }}
\newcommand{\beq}{\begin{equation}}
\newcommand{\eeq}{\end{equation}}

\begin{document}

\thispagestyle{empty}

\begin{flushright}
physics/0106xxx
\end{flushright}

\vskip 2cm

\begin{center}
{\Large\bf  How simple is simple pendulum ?}

\vfill

{\bf  P. Arun  \& Naveen Gaur\footnote{naveen@physics.du.ac.in}}
\\[5mm]   
{Department of Physics \& Astrophysics\\
University of Delhi  \\
Delhi - 110 007\\
India} 
\vfill

\begin{abstract}
The simple pendulum is one of the first experiments that students of
higher physics do. There are certain precautions which the students
are asked to take while performing the experiment. In this note we
will try to explain as to why these precautions are taken up and what
will happen if we relax them.
\end{abstract}	
\end{center}

\vfill
\pagebreak


The simple pendulum is one of the first experiment that physics
students do in higher secondary. Its a fairly simple experiment, which
is then repeated in form of "finding g by bar/ katar's pendulum" in
graduation. Yet a feature noticed is the lack of appreciation by the
students that "the initial angular displacement of the pendulum must
be small", which is disturbing. At the end of the session during viva
examination of the students reveal the ignorance of the students about
this precaution. Answers ranging from "to minimize air resistance" to
"make sure the pendulum won't wobble" are given. In this article we
will approach the problem of convincing students as to why this
precaution is necessary. Also we will try to give some insight of what
could be the possible results if we don't take this precaution. 

\par In the first section we are be going to discuss the simple
pendulum which the student know (where the initial displacement is
small). Here in this section we will set up the relevant equation
first and then solve this equation analytically according to the given
initial conditions. In second section we will set up the equation of
the motion of simple pendulum, when we relax the condition that the
initial displacement from mean position is small. In the third section
we will try to solve the equation which we had set up in second
section. In last section we will derive the conclusions of the whole
exercise. 

\section{Pendulum with small initial displacement}
\begin{figure}[htb]
\epsfig{file=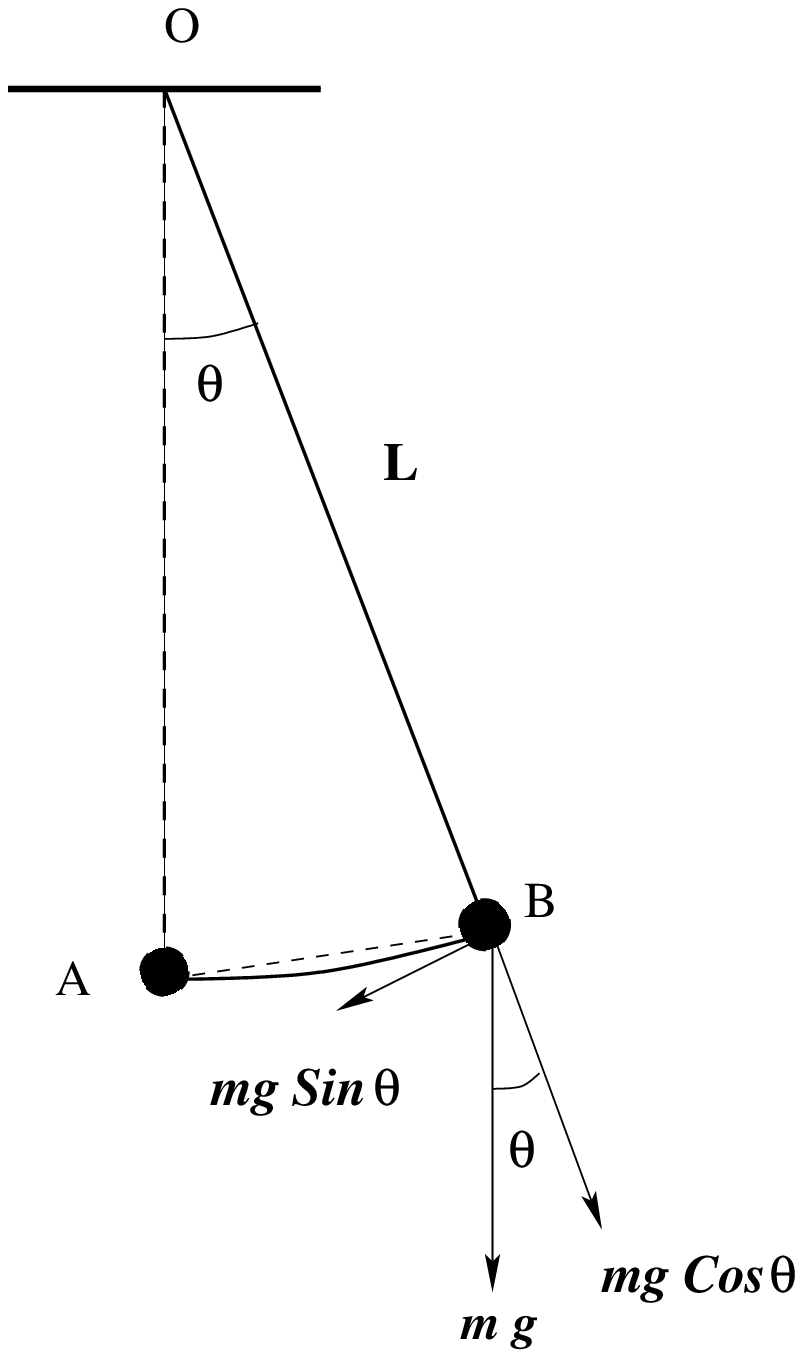,width=3in,height=4in} 
\hfill
\epsfig{file=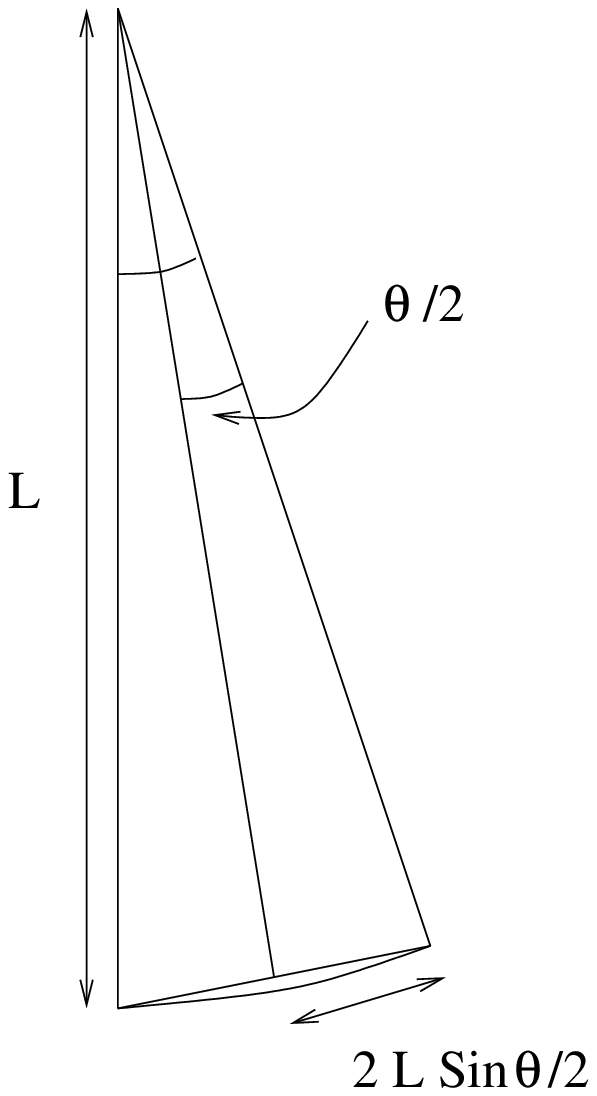,width=3in,height=4in}
\caption{Simple Pendulum of length L}
\vskip -0.3in
\label{fig:1}
\end{figure}

Before proceeding to set up the differential equation of a pendulum
with a large initial displacement from the mean position, first in
this section we see how the usual simple harmonic equation is a result
of the imposed condition that the initial displacement from the mean
position is small. Consider the Figure (\ref{fig:1}), which shows the
bob of the pendulum suspended by a string of length 'L'. The angular
displacement is $\Theta$ and the weight's ($m g$) resolved component
pointing towards the means position is $mg sin\Theta$ . This is the
force, trying to bring back the bob to its mean position, \ie
\beq
F ~=~ -~ m g sin \Theta
\label{eq:1}
\eeq
This force is a restoring force (this is indicated by the negative
sign in eq.(\ref{eq:1})), since it is trying to bring back the
body to its initial (mean) position, since more you move away
from the mean position the larger is the force. That is, the force is
directly proportional to the displacement of of the pendulum from its
mean position. Here the displacement is the length of chord, marked
AB. Since initially, we assume the angular displacement to be very
small, the displacement (length of the chord) and the length of the
arc AB can be assumed equal. Also, as a result of our assumption the
${\rm mgsin\Theta}$ component which acts tangentially at point 'B',
can be assumed to be along the chord. Now the length of chord is
difficult to compute, however, the length of the arc is obtained from
the relation 
\beq
\Theta ~=~ \frac{\rm arc}{\rm radius}
\label{eq:2}
\eeq
Since, the radius of the circle under consideration here is the length
of the thread suspending the bob (with the assumption that the radius
of the bob is insignificant as compared to the length of the thread,
\ie the bob can be considered to be a point mass), the arc length is
given as 
\beq
{\rm arc ~AB} ~=~ L ~\Theta
\label{eq:3}
\eeq
hence, with our assumption that the angular displacement is very
small, the length of the chord (s) is equal to arc AB, i.e. 
\beq
s ~=~ L ~\Theta
\label{eq:4}
\eeq
The equation of motion eq.(\ref{eq:1}), now can be written as
\beq
m~ \frac{d^2s}{dt^2} ~=~ -~m g sin \Theta
\label{eq:5}
\eeq
substituting eqn(\ref{eq:4}) with the knowledge that the length of the
string is constant we have
\beq
\frac{d^2\Theta}{dt^2} ~~=~ - ~\frac{g}{L} ~sin\Theta
\label{eq:6}
\eeq
 In the above equation considering the case of small initial angular
displacement \ie $\Theta$ is small, then ${\rm sin\Theta ~\sim~
\Theta}$ and substituting ${\rm \omega^2 ~=~ g/L}$, we have 
\beq
\frac{d^2\Theta}{dt^2} ~=~ - ~\omega^2 ~\Theta
\label{eq:7}
\eeq
This is the familiar simple harmonic motion (SHM) equation, whose
{\sl general solution} is given by  
\beq
\Theta(t) ~=~ A~ sin(\omega t) ~+~ B~cos(\omega t)
\label{eq:8}
\eeq
where A and B are constants (there would be in general two constants
because we are trying to solve a second order differential
eqn.({\ref{eq:7})). We can get the values of the constants by choosing
suitable initial conditions (in that case we call the solution to be a
{\sl particular solution}). We will discuss this again in last section. 
\par  The time period of oscillation can be obtained from the
relationship ($ \omega ~=~ \frac{2 \pi}{T}$)   
\beq
\omega ~=~ \sqrt{\frac{g}{L}}
\label{eq:9}
\eeq
leading to 
\beq
T~=~2\pi ~\sqrt{L \over g}
\label{eq:10}
\eeq

\section{Pendulum with large initial displacement}

To set up the differential equation for a pendulum which is given a
large displacement, again we have to find the length of the chord
AB. In general, the length of the chord, or the displacement is given
as 
\beq
s ~=~ 2~L~ sin {\Theta \over 2} 
\label{eq:11}
\eeq
considering this as the displacement and using the EOM
(eq.(\ref{eq:1})) 
\begin{eqnarray}
m~ \frac{d^2s}{dt^2} 
  &=&   m~ L~ cos {\Theta \over 2} \frac{d^2\Theta}{dt^2} 
       - \frac{m L}{2} sin{\Theta \over 2} (\frac{d\Theta}{dt})^2
    \nonumber \\ 
  &=& -~ m g ~sin \Theta  
\label{eq:12} 
\end{eqnarray}
so the EOM for large initial displacement is (simplifying
eq.(\ref{eq:12})) 
\beq
cos{\Theta \over 2} ~ \frac{d^2\Theta}{dt^2} ~-~ {1 \over 2}
sin{\Theta \over 2}(\frac{d\Theta}{dt})^2 ~=~ - ~\omega^2 ~sin \Theta  
\label{eq:13}
\eeq
Now our aim is to estimate the error which would be introduced if the
displacement is large. For this we have to solve the above
eqn.({\ref{eq:13}). The term $(d\Theta/dt)^2$ complicates the simple
second order differential equation we have dealt with. Such equations
are called non-linear differential equations and usually are difficult
to solve analytically. Indeed most non-linear equations can only be
solved numerically. We will try to solve this equation numerically in
next section and will try to compare the two solutions.

\section{Numerical results}

\begin{figure}[htb]
\begin{center}
\epsfig{file=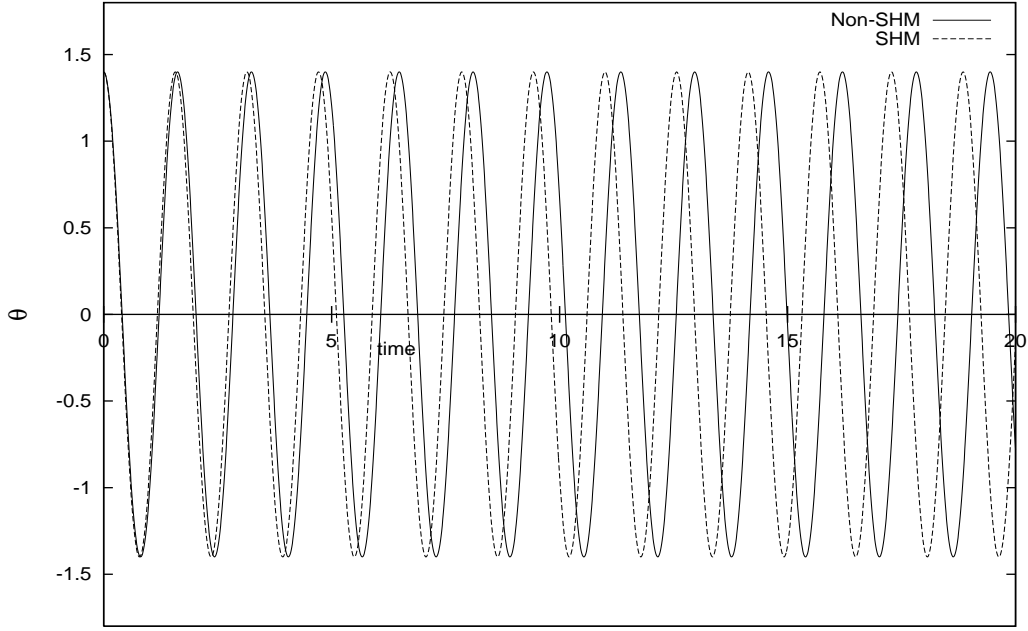,height=3.5in,width=5.5in}
\caption{Plot of $\Theta(t)$ vs t, other parameters are $\omega ~=~ 4
~,~a = 1.4$ (in radians)} 
\label{fig:2}
\end{center}
\end{figure}

As we mentioned in previous section we can't solve the
eqn.(\ref{eq:13}) analytically. So we will try to solve the equation
numerically. Now if we want to solve eqn. numerically we can't get
what is called {\sl general solution}. To get numerical solution we
require the exact numerical values of all the constants and initial
conditions. 

\begin{figure}[htb]
\vskip -0.5in
\begin{center}
\epsfig{file=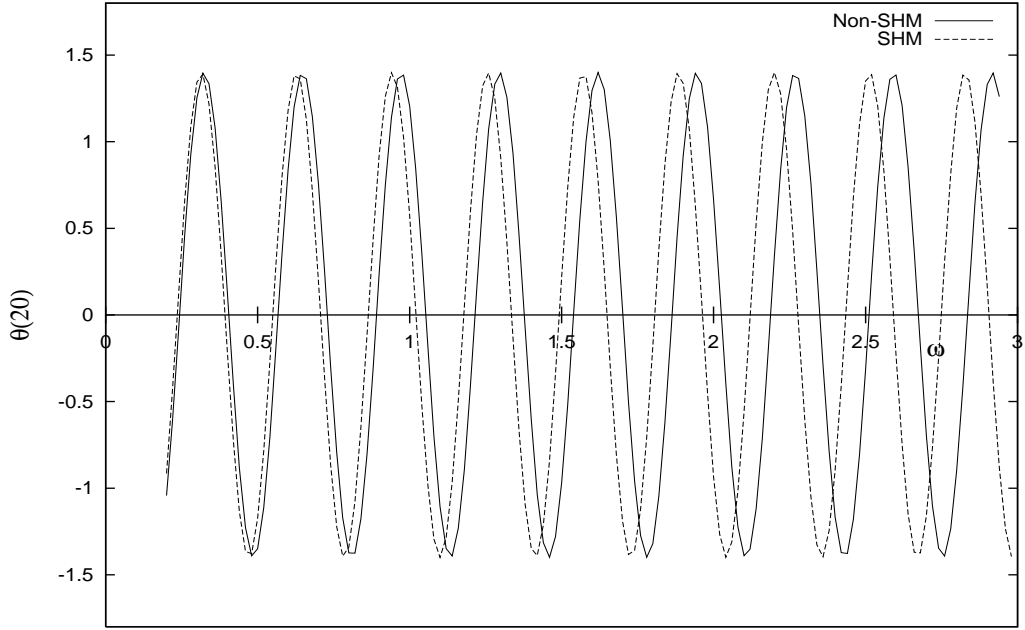,height=3.5in,width=5.5in}
\caption{Plot of $\Theta$ vs $\omega$, other parameters are ${\rm time
~=~20 ~seconds} ~,~a = 1.4$ (in radians)} 
\label{fig:3}
\end{center}
\end{figure}

\par The initial conditions (conditions at $ t = 0$) would be imposed
on $\Theta$ and $d\Theta/dt$. These initial conditions would help us
in finding out the particular solution from the general solution
eq.({\ref{eq:8}}). To solve eq.(\ref{eq:13}) numerically we also
require the value of $\omega$.

\par The initial conditions which we are choosing are 
\beq
\Theta(t = 0) ~=~ a  \quad \quad;\quad \quad 
\frac{d\Theta (t = 0)}{dt} ~=~ 0 
\label{eq:14}
\eeq 
where a is a constant whose value we will take as input. So to
completely solve the equation numerically we have to specify the
values of $a ~{\rm and}~ \omega$.

\begin{figure}[htb]
\vskip -0.5in
\begin{center}
\epsfig{file=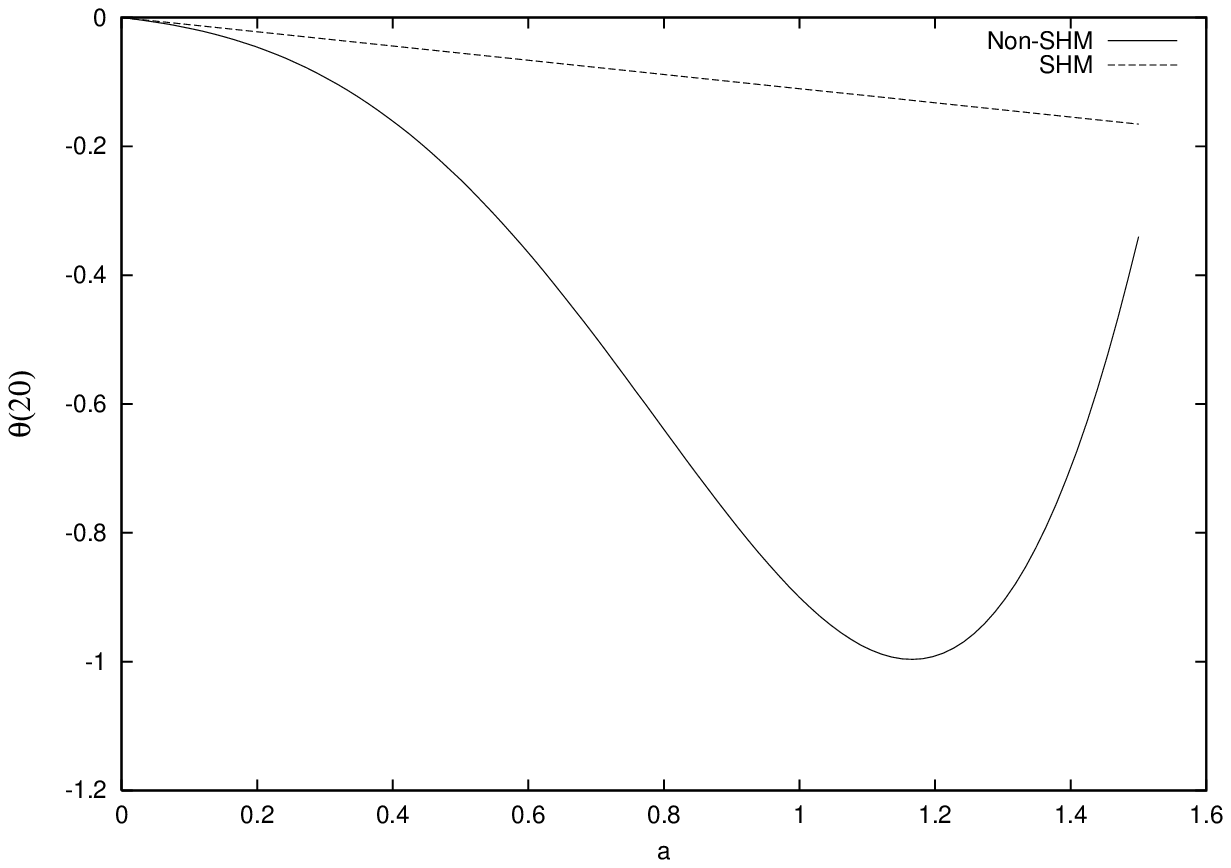,height=3.5in,width=5in}
\caption{Plot of $\Theta$ vs a, other parameters are, ${\rm time ~=~20
~ seconds} ~,~\omega = 4$} 
\label{fig:4}
\end{center}
\end{figure}

Fig(\ref{fig:2}) and Fig(\ref{fig:3}) is a plot exhibiting the
evolution of $\Theta$, the angular  displacement with time and
$\omega$ respectively , based on the results of our numerical
solutions of equation (\ref{eq:7}) and (\ref{eq:13}). For comparing
the possible error that creeps in due to the injudicious choice,
i.e. use eqn(\ref{eq:7}) for large displacement, the numerical
solution of both equations were done for the same initial displacement
a, with the assumption that the ratio of g to L is 16 (\ie $
\omega~=~4$). Remember, small initial displacement condition is 
when you can substitute $\Theta$ instead of $2 sin(\Theta/2)$. It is
evident from fig(\ref{fig:5}), where the straight line shows linear
displacement as a function of angular displacement for small
oscillation condition (eqn (\ref{eq:4})) while actual displacement
follows eqn(\ref{eq:11}). In figure(\ref{fig:5}) we have tried to show
the region (of the angular displacement) in which a motion of pendulum
can be considered to be that of SHM. Keeping this in mind, to emphasis
our point to show errors that are bound to occur, we have done our
calculations with a = 1.4 rad. The variation in the angular
displacement with time, $\Theta(t)$ in fig(\ref{fig:2}), for the two
cases are very different, especially with increasing time. This
indicates disparity between the two cases. It would hence, be 
interesting to note that initially there is not much difference in
position at a given instant between the two cases, as can be
seen. However, as time progresses disparity increases. This disparity
is found to depend on the size of the initial displacement, as shown
in Fig(\ref{fig:4}). Fig(\ref{fig:4}) is the angular displacement of 
pendulum after 40sec of the sustained oscillations for increasing
initial displacement. The choice is just to show disparity between the
two cases soon after oscillations have commenced. Figure (\ref{fig:4})
shows how much the disparity between the two cases increases with
increasing initial angular displacement, in turn the deviation from
"small displacement condition". 

\begin{figure}[htb]
\vskip -0.5in
\begin{center}
\epsfig{file=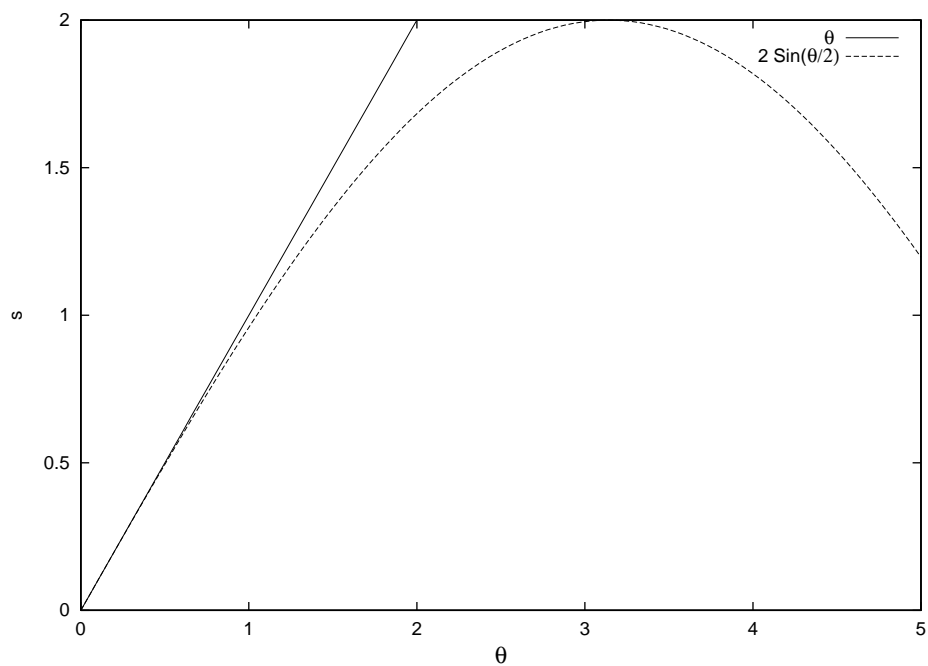,height=3.5in,width=5in}
\caption{Plot of linear displacement with angular displacement} 
\label{fig:5}
\end{center}
\end{figure}
\begin{figure}[htb]
\vskip -0.5in
\begin{center}
\epsfig{file=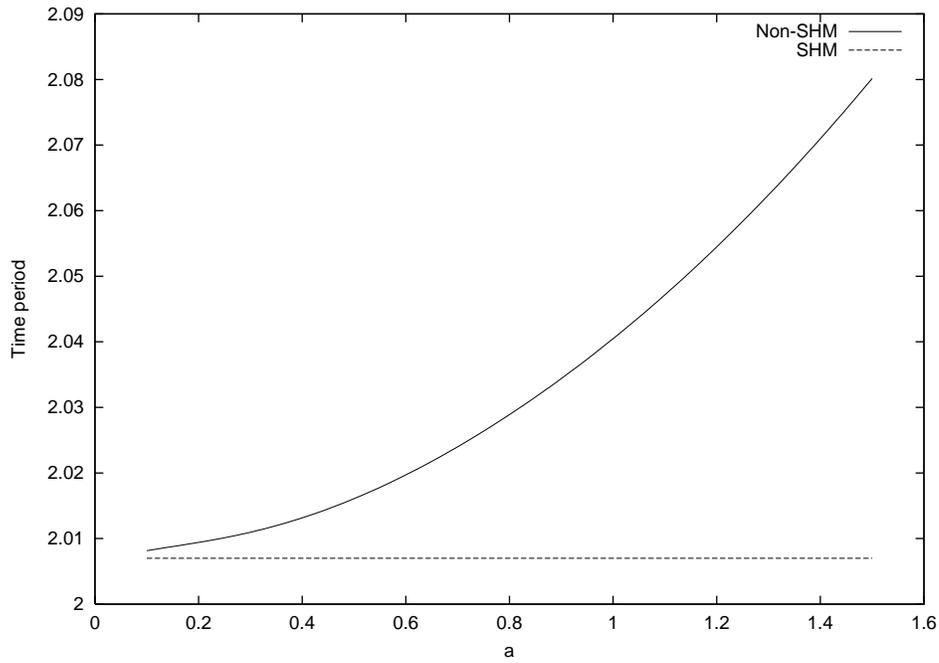,height=3.5in,width=5in}
\caption{Plot of Time period with the initial displacement (in
radians), other parameter is $\omega ( = 3.13)$} 
\label{fig:6}
\end{center}
\end{figure}
\begin{figure}[htb]
\vskip -0.5in
\begin{center}
\epsfig{file=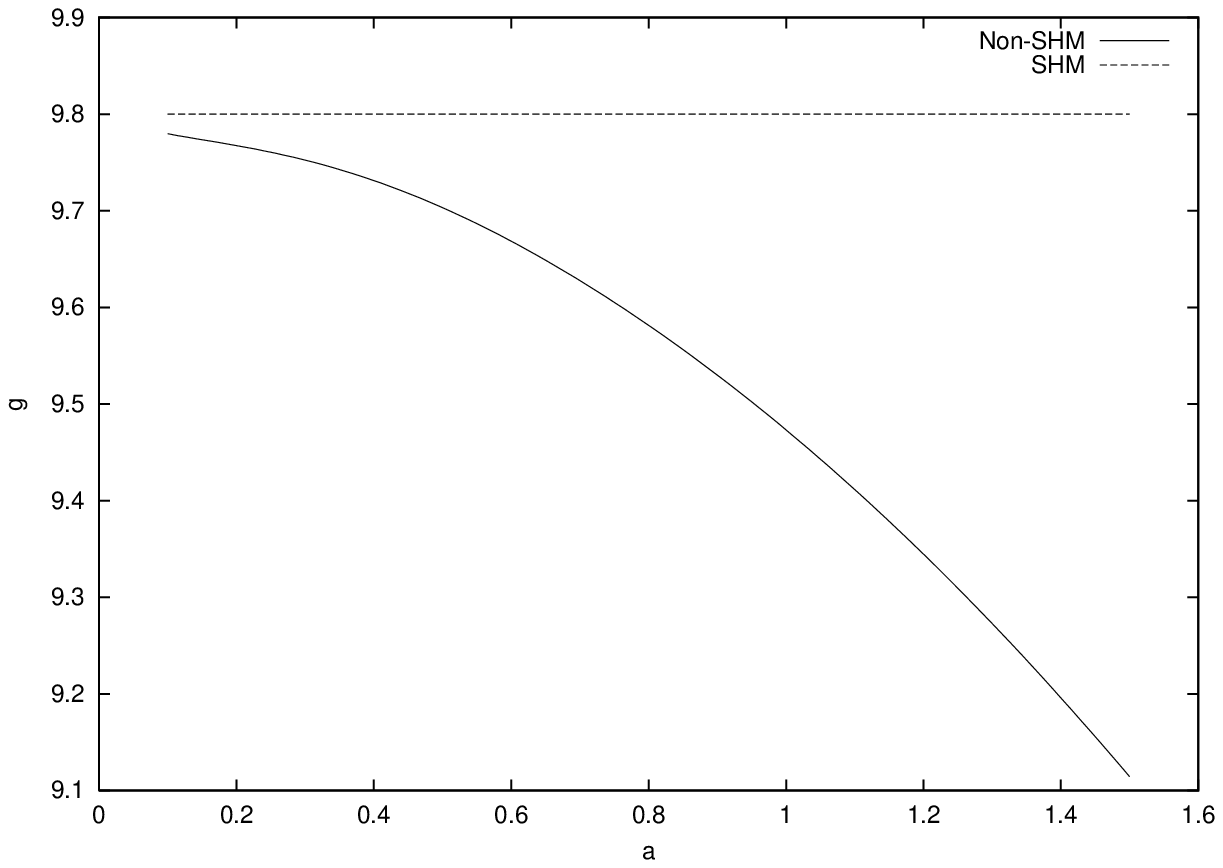,height=3.5in,width=5in}
\caption{Plot of acceleration due to gravity with initial displacement
other parameters are same as that of Figure(\ref{fig:6})}
\label{fig:7}
\end{center}
\end{figure}

\section{Conclusion}

Now let's summarize the conclusions of the whole exercise. We know
that only small initial displacement can exhibit SHM motion, while
large initial displacement will execute Non SHM. A simple pendulum
experiment is usually done to establish the value of the acceleration
due to gravity. To appreciate the error that will occur if SHM
equations are used while actually the initial displacement warrants
the use of the non-linear equations, we calculate the average time
period with large displacements using the above data. For SHM case the
time period would be a constant for any value of initial
displacement. For large initial displacement as we can see from
Figure(\ref{fig:2}) the motion of pendulum won't be
simple harmonic. Typically a student find out time period by taking the
time taken for a certain no. of oscillations, then dividing this time
by no. of oscillations. For Figure(\ref{fig:6}) we have numerically
found the time period by calculating time taken for 23
oscillations. In last graph (Figure(\ref{fig:7})) we have shown the
plot of the variation of acceleration due to gravity with the initial
displacement based on the time periods thus calculated from
fig(\ref{fig:6}) and substituting in eqn(\ref{eq:10}). As we can see
for SHM motion this would be a straight line. But once one consider
large initial displacement this would no longer be the case, and the
variation the student can expect in his exercise could be fairly
large. 

\par In summary we have tried to show the importance of the condition
of small initial displacement in the whole exercise of performing a
experiment of calculation of acceleration due to gravity by simple
pendulum.

\end{document}